\documentclass[modern]{aastex631}

\received{\today}
\revised{\today}
\accepted{\today}

\newcommand{\teff}{$T_\mathrm{eff}$}
\newcommand{\mjup}{$M_\mathrm{Jup}$}
\newcommand{\msun}{$M_\mathrm{\odot}$}
\newcommand{\rsun}{$R_\mathrm{\odot}$}
\newcommand{\lsun}{$L_\mathrm{\odot}$}
\newcommand{\orvara}{\texttt{orvara}}

\newcommand{\age}{$4.6^{+3.8}_{-1.3}$}

\newcommand{\bmass}{${9.1}_{-1.1}^{+1.0}$}
\newcommand{\binc}{${32.7}_{-3.2}^{+5.3}$}
\newcommand{\becc}{${0.3686}_{-0.0031}^{+0.0032}$}
\newcommand{\bsma}{${2.845}_{-0.039}^{+0.038}$}
\newcommand{\bper}{${4.8277}_{-0.0023}^{+0.0022}$}

\newcommand{\cmass}{${6.9}_{-1.0}^{+1.7}$}
\newcommand{\cinc}{${101}_{-33}^{+31}$}
\newcommand{\cecc}{${0.64}_{-0.13}^{+0.12}$}
\newcommand{\csma}{${27.4}_{-7.9}^{+16}$}
\newcommand{\cper}{${144}_{-58}^{+139}$}

\newcommand{\thetabc}{$96.3_{-36.8}^{+29.1}$}
\newcommand{\rvoff}{$-5.3_{-2.0}^{+2.0}$}
\shorttitle{The misaligned planets of 14 Her}
\shortauthors{Bardalez Gagliuffi et al.}

\usepackage{graphics}
\usepackage{hyperref}

\begin{document}

\title{14 Her: a likely case of planet-planet scattering}

\correspondingauthor{Daniella C. Bardalez Gagliuffi}
\email{dbardalezgagliuffi@amnh.org}

\author[0000-0001-8170-7072]{Daniella C. Bardalez Gagliuffi}
\affiliation{American Museum of Natural History, 200 Central Park West, New York, NY 10024, USA}

\author[0000-0001-6251-0573]{Jacqueline K. Faherty}
\affiliation{American Museum of Natural History, 200 Central Park West, New York, NY 10024, USA}

\author[0000-0002-6845-9702]{Yiting Li}
\affiliation{Department of Physics, University of California, Santa Barbara, Santa Barbara, CA 93106, USA}

\author[0000-0003-2630-8073]{Timothy D.\ Brandt}
\affiliation{Department of Physics, University of California, Santa Barbara, Santa Barbara, CA 93106, USA}

\author[0000-0002-0604-5440]{Lauryn Williams}
\affiliation{Department of Physics \& Astronomy, University of Missouri - Columbia, Columbia, MO 65202, USA}

\author[0000-0003-0168-3010]{G.\ Mirek Brandt}
\affiliation{Department of Physics, University of California, Santa Barbara, Santa Barbara, CA 93106, USA}

\author{Christopher R. Gelino}
\affiliation{Caltech/IPAC-NExScI, Mail Code 100-22, Caltech, 1200 East California Blvd., Pasadena, CA 91125, USA}

\begin{abstract}
In this Letter, we measure the full orbital architecture of the two-planet system around the nearby K0 dwarf 14~Herculis.~\objectname{14 Her} (HD 145675, HIP 79248) is a middle-aged (\age\,Gyr) K0 star with two eccentric giant planets identified in the literature from radial velocity (RV) variability and long-term trends. Using archival RV data from Keck/HIRES in concert with \textit{Gaia-Hipparcos} acceleration in the proper motion vector for the star, we have disentangled the mass and inclination of the b planet to \bmass\,\mjup~and \binc\,degrees. Despite only partial phase coverage for the c planet's orbit, we are able to constrain its mass and orbital parameters as well to \cmass\,\mjup~and \cinc\,degrees. We find that coplanarity of the b and c orbits is strongly disfavored. Combined with the age of the system and the comparable masses of its planets, this suggests that planet-planet scattering may be responsible for the current configuration of the system.
\end{abstract}


\vspace{-2.7cm}

\section{Introduction} \label{sec:intro}

The orbital parameters of a planetary system are sculpted by its formation processes and evolutionary history in a quest for a stable configuration. Circular, coplanar planetary orbits are a natural consequence of formation in a disk, whether by core accretion or gravitational instability~\citep[e.g.,][]{1993ARAandA..31..129L,2001ASPC..235..195B}. This effect is observed in the low eccentricities and low mutual inclinations of the Solar System planets, and multi-exoplanet systems~\citep[e.g.,][]{2015PNAS..112...20L}. ALMA observations of protoplanetary disks with gaps also suggest coplanarity likely as a result of newborn planets clearing gas and dust from the disk~\citep[e.g., HL Tau][]{2015MNRAS.453L..73D}.

The dynamical evolution of an early planetary system determines the final configuration of its components \citep{2008ApJ...686..580C,2019A&A...629L...7C,2008ApJ...686..621F}. Fly-by events can excite orbital eccentricities and even eject planets on timescales proportional to the impact parameter of the fly-by~\citep[e.g.,][]{2011MNRAS.411..859M}. Fly-bys can also trigger planet-planet scattering, decreasing the semimajor axes of some planets (typically the more massive ones) while leaving others in much wider orbits~\citep{2009ApJ...693L.113S,2009ApJ...696.1600V}.
The orbital architectures of mature systems reflect both their formation conditions and their dynamical evolution.


In this Letter, we present the full orbital architecture of the 14~Her system.  
Over 20 years of RV follow-up show a clear signature of a giant planet~\citep[e.g.][]{2003ApJ...582..455B,2004A&A...414..351N,2006ApJ...645..688G}, and a long-term trend for a second one~\citep[e.g.,][]{2021AJ....161..134H,2021ApJS..255....8R,2007ApJ...654..625W}. With absolute astrometry from \emph{Hipparcos} and \emph{Gaia}, 
we are able to break the $M \sin i$ degeneracy for both planets and calculate their orbital parameters, as well as identify strong evidence for a high mutual inclination between the orbits. The orbits of the 14~Her planets hint at a turbulent past marked by dynamical interactions between the two massive planets that led to their current, peculiar configurations.


We structure the Letter as follows.  In Section~\ref{sec:starchar}, we present the existing stellar characteristics from the literature. In Section~\ref{sec:data}, we describe our RV and absolute astrometric data.  Section~\ref{sec:mcmc} describes the \orvara~tool for orbit fitting and our analysis procedure. In Section~\ref{sec:results}, we report the orbital parameters derived for the two planets of this system. In Section~\ref{sec:discussion}, we explore the implications of our results for the formation and evolution of the 14~Her system. We present our conclusions in Section~\ref{sec:conclusions}.

\section{System Characterization}\label{sec:starchar}

14~Her ($\alpha$ = 16:10:24.315, $\delta$ = +43:49:03.50) is a middle-aged K0 dwarf located $17.9416\pm0.0072$\,pc away with an estimated \teff $= 5282$\,K~\citep{2021AandA...649A...1G,2021A&A...649A...2L}. 
Due to its brightness and proximity to Earth, 14~Her was one of the first stars monitored with RV to search for exoplanets~\citep[e.g.,][]{2003A&A...410.1039P,2003A&A...410.1051N,2004A&A...414..351N}. 
Even though significant RV variations had been found in the ELODIE RV data by 1996, the first formal discovery publication for 14~Her~b was not presented until 2003~\citep{2003ApJ...582..455B}. 
14~Her~c was not identified until 2007~\citep{2007ApJ...654..625W} due to its long orbital period.

Similarly to other planet-hosting stars discovered around the same time (e.g., $\rho^1$ Cnc), 14~Her is ``super''-metal-rich with a $\mathrm{[Fe/H]} = 0.50\pm0.05$~\citep{1996ApJS..102..105T,1999ApJ...511L.111G}. Alternative measurements place the metallicity of 14~Her between $\mathrm{[Fe/H]} = 0.30-0.60$~\citep[e.g.,][]{2003AJ....126.2015H,2006AJ....131.3069L}, although certainly metal-rich. Abundance measurements of absorption lines show that 14~Her is a chromospherically quiet star~($log_{10} R'_{HK} = -4.94\pm0.04$;~\citealt{2019AJ....158..101M}).


We infer the fundamental parameters of 14~Her using the Bayesian activity-age dating method of \cite{2014ApJ...786....1B}, the luminosity, effective temperature, and angular diameter relations of \cite{2010A&A...512A..54C}, and the PARSEC isochrones \citep{2012MNRAS.427..127B}.  We find an age of $4.6^{+3.8}_{-1.3}$\,Gyr based on the star's low chromospheric activity and slow rotation period.  We use the $V_T-J$ color from Tycho-2 \citep{2000A&A...355L..27H} and 2MASS \citep{2003yCat.2246....0C} and adopt a metallicity of $[{\rm Fe/H}] = 0.43 \pm 0.07$ to infer a luminosity of $0.67 \pm 0.02$\,$L_\odot$ using the relations given in \cite{2010A&A...512A..54C}.  Our adopted metallicity range spans 2/3 of the measurements in the PASTEL catalog \citep{2010AandA...515A.111S}, all of which are from high-dispersion, high signal-to-noise spectroscopy.  We do require a slight extrapolation of the \cite{2010A&A...512A..54C} relations, which are only validated to $[{\rm Fe/H}] = 0.4$.  The effective temperature relations of \cite{2010A&A...512A..54C}, based on the $V_T - K_s$ color, give $T_{\rm eff} = 5310 \pm 30$\,K after including both measurement errors and the 18\,K rms scatter about their calibrated relation.  These combine with the measured \emph{Gaia} parallax to give a radius of $0.97 \pm 0.02$\,$R_\odot$.  This compares well with the $J$-band-based angular radius of $0.99 \pm 0.02$\,$R_\odot$ using the $V_T-K_s$ color \citep[Table 6,][]{2010A&A...512A..54C}.  Finally, we use our inferred age distribution, a uniform $[{\rm Fe/H}]$ distribution, and the Salpeter initial mass function as priors, and construct a posterior mass distribution using the PARSEC isochrones together with our inferred metallicity and luminosity.  We finally obtain a mass for 14~Her~A of $0.98 \pm 0.04$\,$M_\odot$.

\begin{figure}
    \centering
    \includegraphics[width=0.77\textwidth]{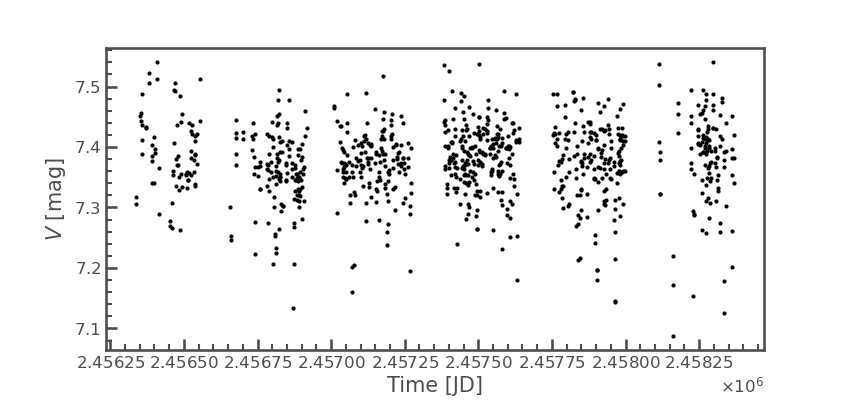}\\
    \includegraphics[width=0.77\textwidth]{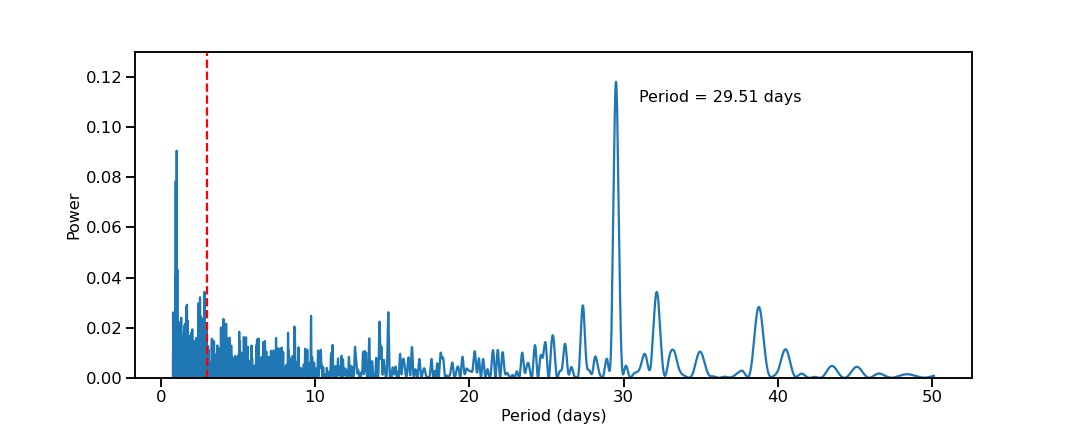}
    \caption{\emph{(Top)} ASAS-SN light curve for 14~Her. \emph{(Bottom)} Lomb-Scargle periodogram showing a significant period at 29.5\,days. Periods shorter than the red line are shorter than the 2-3\,day ASAS-SN cadence.}
    \label{fig:14HerLC}
\end{figure}

Previous estimations of the rotation period of this star place it at 22.38\,days~\citep{2016AandA...596A..76S} and $48.5\pm1.137$\,days~\citep{2010MNRAS.408.1606W}, both of these coming from activity-period relations based on $logR'_{HK}$ measurements. While 14~Her has been observed in \emph{TESS} sectors 24 and 25, we are limited by the observing window of a given sector of 27.4\,days, and we do not attempt to combine \emph{TESS} sectors since is not trivial. From an archival ASAS-SN light curve, spanning just over 5 years (2013-02-13 to 2018-09-06), we have obtained a significant period of 29.5\,days with a Lomb-Scargle periodogram (Figure~\ref{fig:14HerLC}), which is consistent with the age of the star.

\begin{deluxetable*}{lcc}
\tablenum{1}
\tablecaption{Stellar properties.}\label{tab:star}
\tablewidth{0pt}
\tablehead{
\colhead{Property} & \colhead{Value} & \colhead{Ref.}}
\startdata
\hline
\multicolumn{3}{c}{\emph{Fundamental Parameters}} \\
Spectral type & K0V & 4 \\
Effective temperature (\teff) & $5310\pm30$\,K & 1\\
Mass (M) & $0.98\pm0.04$\,\msun & 1 \\
Age & $4.6^{+3.8}_{-1.3}$\,Gyr & 1\\
Radius (R) & $0.97 \pm 0.02\,R_{\odot}$ & 1\\
Luminosity (L) & $0.67\pm0.02\,$\lsun & 1 \\
Surface gravity (log $g$) & 4.46 & 3\\
Bulk metallicity ($\mathrm{[Fe/H]}$) & $0.43\pm0.07$ & 5\\
Chromospheric activity ($log R'_{HK}$) & $-4.94\pm0.04$ & 3\\
Equatorial velocity ($V\,\sin i$) & $1.65$\,km/s & 3\\
Rotation period (P$_{rot}$) &  29.5\,days & 1\\
\hline
\multicolumn{3}{c}{\emph{Astrometry}} \\
Parallax & $55.866\pm0.029$\,mas & 2\\
Proper Motion in R.A. ($\mu_{\alpha}cos\delta$) & $131.745\pm0.028$\,mas/yr & 2\\
Proper Motion in Dec. ($\mu_{\delta}$) & $-297.025\pm0.037$\,mas/yr & 2\\
$\chi^2$ & 1008.64 & 1\\
\hline
\multicolumn{3}{c}{\emph{Photometry}} \\
\emph{Gaia} $B_p-R_p$ color & 1.002\,mag & 2\\
\emph{Gaia} $G$ magnitude & 6.395\,mag & 2\\
\emph{Gaia} RUWE & 1.819 & 2
\enddata
\tablerefs{(1) This paper; (2)~\citet{2021AandA...649A...1G}; (3)~\citet{2019AJ....158..101M}; (4)~\citet{2018ApJS..238...29P}; (5)~\citet{2010AandA...515A.111S}.}
\end{deluxetable*}

\section{Data}\label{sec:data}

\subsection{Keck/HIRES radial velocity}

\begin{figure}
    \centering
    \includegraphics[width=0.77\textwidth]{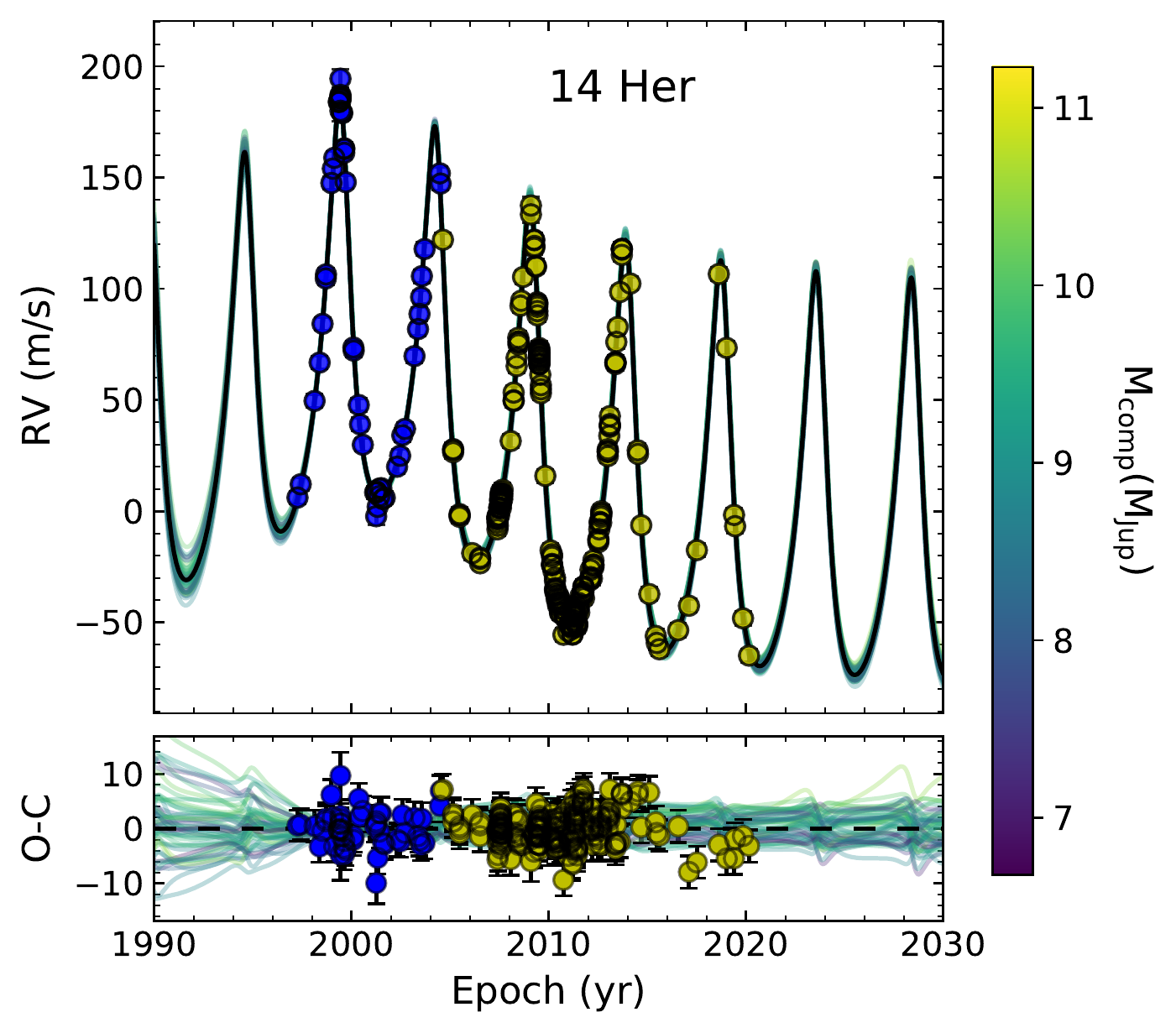}
    \caption{\orvara~fit to the HIRES RV curve spanning over 20 years. Data from the pre-2004 and post-2004 upgrade are shown as blue and yellow solid circles, respectively. Residuals are show in the bottom panel. Colorbar shows the range of masses of the b planet from the MCMC chains.}
    \label{fig:rv}
\end{figure}

We have collected 283 archival RV data points from Keck/HIRES~\citep{2021ApJS..255....8R,2003ApJ...582..455B,2007ApJ...654..625W,2021AJ....161..134H} spanning over 20 years (1997-04-07 to 2020-02-26;~\citealt{2021ApJS..255....8R}). The High Resolution Echelle Spectrometer (HIRES;~\citealt{1994SPIE.2198..362V}), mounted on the Keck I 10-m Telescope, operates between $0.3-1\,\mu$m providing spectral resolutions between R$\sim25000-85000$ and RV precision down to 1\,m/s. The RV measurements for 14~Her fluctuate between $-$73\,m/s and 191\,m/s, with a semi-amplitude of 100\,m/s and a visible long-term slope caused by the c planet. The mean precision for our data is 1.08\,m/s. The RV curve of 14~Her is shown in Figure~\ref{fig:rv}.

We also found RV data in the literature from the ELODIE spectrograph at the Observatoire de Haute-Provence and the Automated Planet Finder (APF) at Lick Observatory. The ELODIE data spans epochs from 1997 to 2002, whereas the APF data also covers 5 years but between 2014 and 2019. The epoch span of the ELODIE data was sufficient to discover the b planet and fully determine its orbit~\citep{2004A&A...414..351N}. However, we require a longer baseline to determine the orbit of the c planet, and HIRES is the longest-running instrument with continuous RV monitoring of planet-hosting stars~\citep{2017AJ....153..208B}. Therefore, we decided to only use HIRES RV data to avoid any offsets or potential systematics across instruments. However, HIRES underwent an upgrade in 2004 which effectively changed its RV zero-point (~\citealt{2019MNRAS.484L...8T}; see Section~\ref{sec:rvoff}).We conservatively treat the pre- and post-upgrade data as coming from two different instruments which introduces two more degrees of freedom to the orbit fit.

\subsection{\textit{Hipparcos} and \textit{Gaia} absolute stellar astrometry}

We combine the radial acceleration information with tangential acceleration from absolute astrometry. The long baseline between \emph{Hipparcos} and \emph{Gaia} epochs and increasingly better astrometric precision from \emph{Gaia} EDR3 significantly improve the uncertainty of the proper motion accelerations. We use the absolute astrometry of the Hipparcos-Gaia Catalog of Accelerations (HGCA;~\citealt{2021ApJS..254...42B}) which has now been updated with \emph{Gaia} EDR3 astrometry. The HGCA has cross-calibrated the astrometric solutions for all \emph{Hipparcos} stars that are present in \emph{Gaia} to a common reference frame~\citep{2018ApJS..239...31B}.

The proper motion of 14~Her across the \emph{Hipparcos} and \emph{Gaia} EDR3 epochs is inconsistent with a constant value by $31\sigma$. 14~Her has a mean acceleration in its proper motion direction of $ (a_{\alpha},a_{\delta}) = (0.812,-8.918)$\,m/s/yr. Its total acceleration ($a_{tot} = 8.955$\,m/s/yr) is lower than that for 97\% of stars in the HGCA, therefore it is a highly significant, low accelerating star.

\section{Three-body orbit fit with \textit{orvara}\label{sec:mcmc}}

In order to fully determine the orbits of the 14~Her planets, we use the RV and absolute astrometry data in concert. The Orbits from Radial Velocity, Absolute, and/or Relative Astrometry Python package (\orvara;~\citealt{2021arXiv210511671B}) works by fitting Keplerian orbits to a combination of absolute astrometry from \emph{Hipparcos-Gaia}, RV, and relative astrometry if available. 
For 3-body orbits, as is the case for 14~Her, the star's motion is approximated as a superposition of one Keplerian orbit due to each companion (see \citealt{2021AJ....161..179B} for a demonstration and validation of the approach). 

We use \orvara~to infer masses and orbital parameters for both 14~Her planets simultaneously. We use a parallel-tempered MCMC sampler \citep{2013PASP..125..306F,2021ascl.soft01006V} to robustly explore the parameter space with several copies of the system, randomly initialized at 30 temperatures. For each temperature, we use 100 walkers with 600,000 steps per walker, discarding the first 30,000 steps as burn-in and saving every $50^{th}$ step which are then used for inference. 
For our final values, we combined four such chains into a long pseudo-chain. We set a Gaussian prior on the primary mass of $0.98\pm0.04$\,\msun and a log-flat prior on the RV jitter to range between $10^{-5}-10$\,m/s, letting the jitter of the pre- and post-2004 HIRES upgrade vary independently. We test the convergence of the MCMC algorithm by running several 600,000 step chains and confirming a stable plateau in log likelihood for both planet solutions.

\section{Results\label{sec:results}}

\subsection{Orbital parameters of 14~Her~b and c}

\begin{figure}
    \centering
    \includegraphics[width=0.77\textwidth]{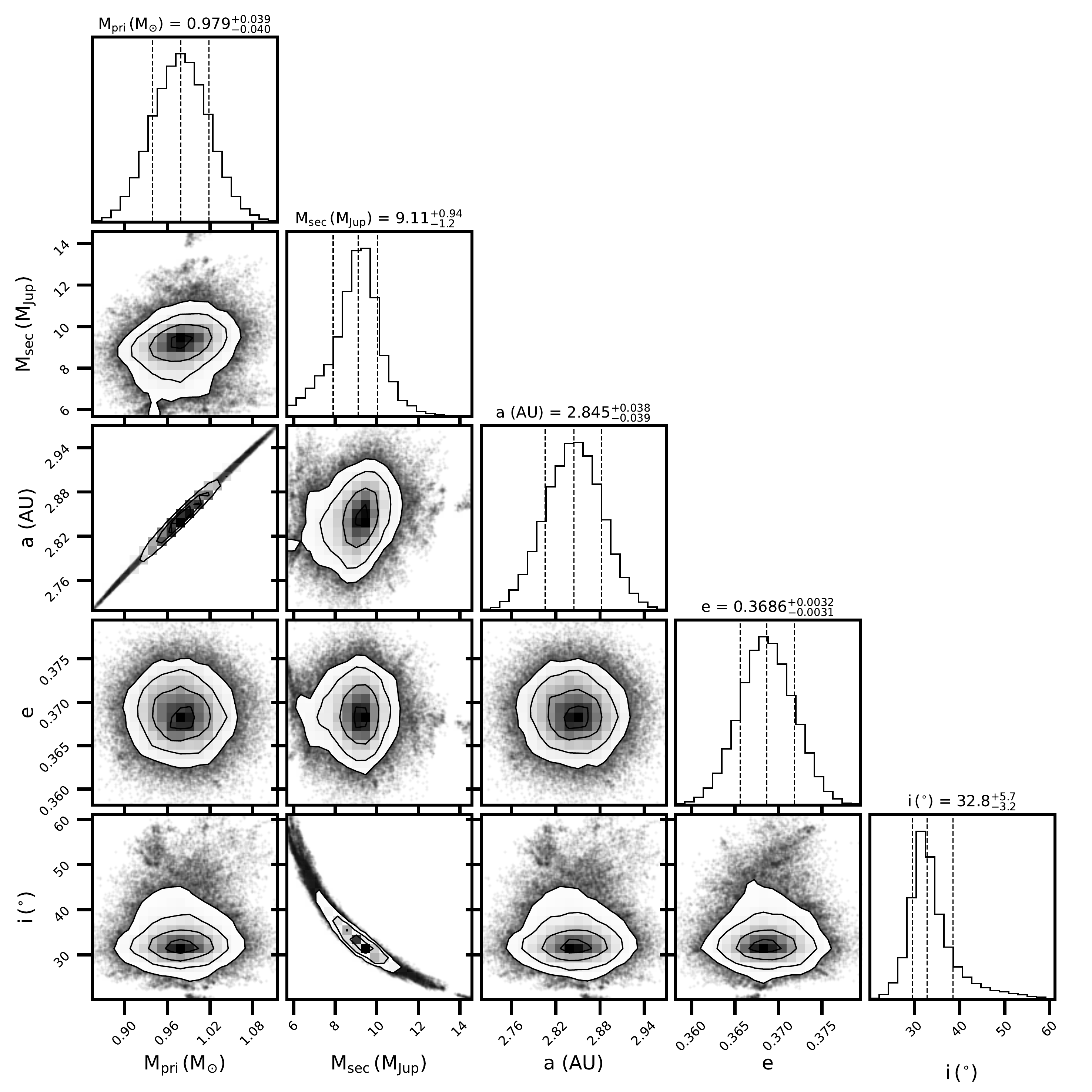}
    \caption{Corner plot for the orbital parameters of the b planet.}
    \label{fig:bcorner}
\end{figure}

Our derived stellar and planetary parameters are shown in Table~\ref{tab:props}. 
Since our RV data covers nearly five full orbits of the b planet, we are able to set strong constraints on its orbital parameters (Figure~\ref{fig:bcorner}). The b planet has a mass of M$_b =\,$\bmass\mjup~and its orbit has a semimajor axis of $a = $\bsma\,AU with a moderate eccentricity of $e = $\becc and a \binc\,degree inclination with respect to the plane of the sky. Its orbital period translates to $P = $\bper\,years. 

\begin{figure}
    \centering
    \includegraphics[width=0.77\textwidth]{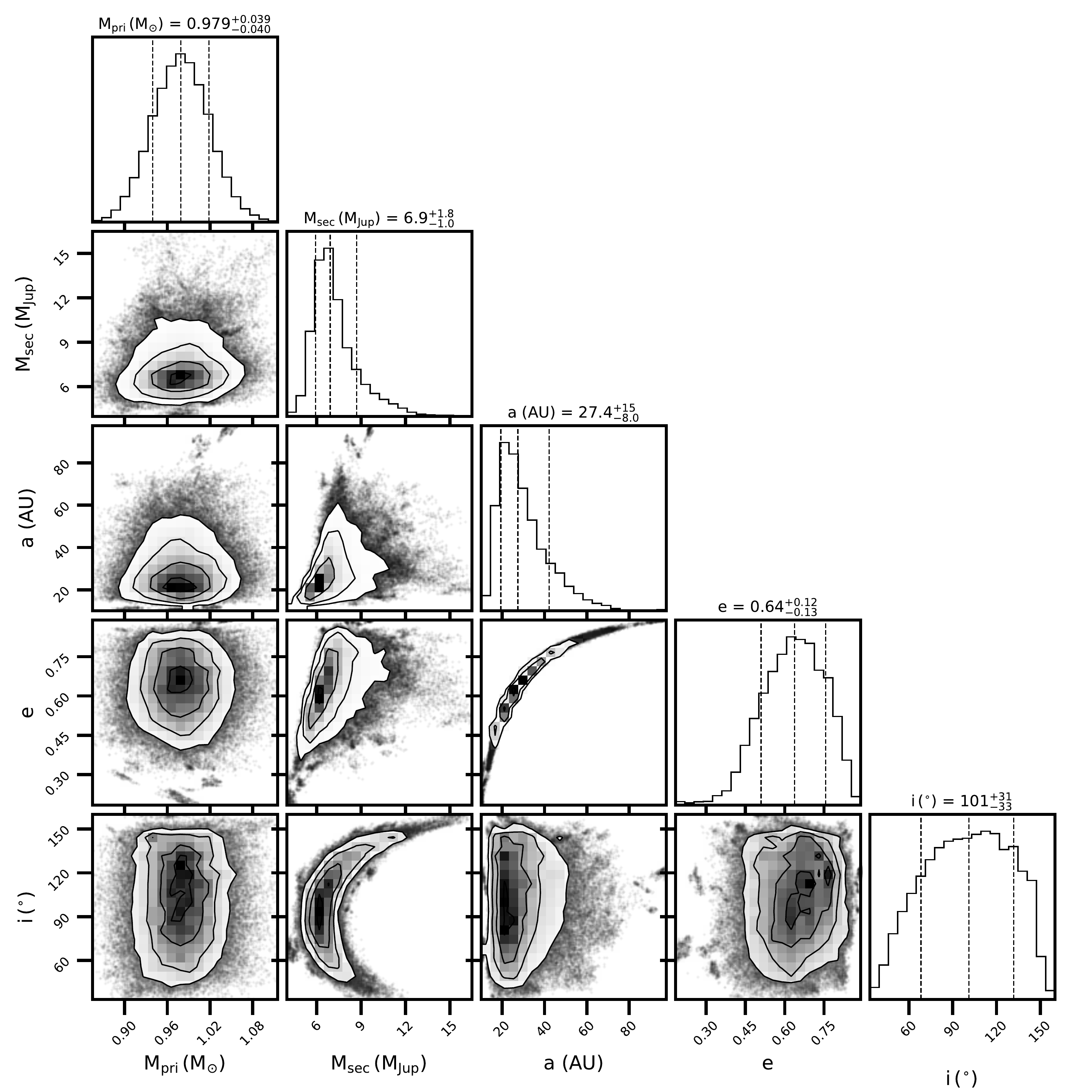}
    \caption{Corner plot for the orbital parameters of the c planet.}
    \label{fig:ccorner}
\end{figure}

Despite only seeing a long-term trend in the RV data covering $\sim$15\% of the period, we can set some constraints on the orbital parameters of the c planet. The mass of the c planet is lower than that of b (M$_c =\,$\cmass\,\mjup) and it is located much farther away from the star, at  $a = $\csma\,AU. 14~Her~c has a highly eccentric orbit ($e =$\cecc). With respect to the orbit of 14~Her~b, the c planet has a broad distribution of inclinations ($i =$\cinc\,degree) but hints at being misaligned (see Figure~\ref{fig:ccorner}). Its orbital period amounts to $P =$\cper\,years with large uncertainties. Our $M\sin i$, semimajor axis, and eccentricity values for the b planet are in line with previous results~\citep[e.g.,][]{2021AJ....161..134H,2021ApJS..255....8R}. However, introducing the constraint from the absolute astrometry enhances these parameters for the c planet with respect to earlier determinations using only RV data. We calculate a minimum mass of $M_c\,\sin i_c$ = $6.79^{+1.85}_{-1.03}$\,\mjup~which is higher but consistent within $1\sigma$ with published values ($M_c\,\sin i_c$ = $5.8^{+1.4}_{-1.0}$\,\mjup,~\citealt{2021ApJS..255....8R}; $M_c\,\sin i_c$ = $6.1^{+1.3}_{-0.9}$,~\citealt{2021AJ....161..134H}). 

\begin{deluxetable*}{lcc}
\tablenum{2}
\tablecaption{MCMC-derived properties for the 14 Her planets. \label{tab:props}}
\tablewidth{0pt}
\tablehead{\colhead{Parameter} & \multicolumn{2}{c}{Posterior Median $\pm1\sigma$}}
\startdata
\multicolumn{3}{c}{\emph{System parameters}} \\
 & \multicolumn{2}{c}{\emph{14 Her}} \\
 \hline
Stellar mass ($M_{*}$) & \multicolumn{2}{c}{${0.984}_{-0.046}^{+0.047}$\,\msun} \\
Parallax ($\varpi$) & \multicolumn{2}{c}{${55.86573}_{-0.00049}^{+0.00046}$\,mas} \\
RV jitter ($\sigma_{jit}$, pre-upgrade) & \multicolumn{2}{c}{${2.85}_{-0.36}^{+0.42}$\,m/s} \\
RV jitter ($\sigma_{jit}$, post-upgrade) & \multicolumn{2}{c}{${2.88}_{-0.15}^{+0.16}$\,m/s} \\
RV zero-point (pre-upgrade) &  \multicolumn{2}{c}{${6.9}_{-8.6}^{+9.1}$\,m/s} \\
RV zero-point (post-upgrade) & \multicolumn{2}{c}{${12.1}_{-8.1}^{+8.9}$\,m/s} \\
\orvara\,reference epoch ($t_{\mathrm{ref}}$) & \multicolumn{2}{c}{2455197.50\,BJD} \\
\hline
\multicolumn{3}{c}{\emph{Planetary parameters}} \\
 & $b$ & $c$ \\
 \hline
 Planet mass ($M$) & \bmass\,\mjup & \cmass\,\mjup\\
 Mass ratio (q) &  ${0.00892}_{-0.0011}^{+0.00090}$ & ${0.00674}_{-0.00095}^{+0.0017}$ \\
 Semi-major axis ($a$) & \bsma\,AU & \csma\,AU \\
 Semi-major axis ($\alpha$) & ${158.9}_{-2.2}^{+2.1}$\,mas & ${1529}_{-442}^{+869}$\,mas \\
 Period (P) & \bper\,yrs & \cper\,yrs \\
 Eccentricity ($e$) & \becc & \cecc \\
 Inclination ($i$) & \binc\,deg & \cinc\,deg\\
 Argument of periastron ($\omega$) &  ${22.78}_{-0.55}^{+0.53}$\,deg & ${15.2}_{-6.0}^{+6.0}$\,deg \\
 PA of ascending node ($\Omega$) & ${236}_{-15}^{+15}$\,deg & ${313}_{-57}^{+30}$\,deg \\
 Mean longitude at reference epoch  & ${82.71}_{-0.19}^{+0.19}$ & ${36}_{-10}^{+12}$\,deg \\
 Epoch at periastron (T$_0$) & ${2456667.4}_{-2.2}^{+2.3}$\,JD & ${2504873}_{-21163}^{+50765}$\,JD\\
 Minimum mass ($M\,\sin i$) &  $4.93^{+0.51}_{-0.68}$\,\mjup & $6.79^{+1.85}_{-1.03}$\,\mjup 
\enddata
\end{deluxetable*}

\subsection{Relative orientation of 14~Her~b and c\label{sec:imut}}

A surprising aspect of our MCMC results is the evidence for discrepant inclinations of the orbital planes for the two 14~Her planets. Orbit misalignment of outer giant planets can have an effect on the final configuration of super-Earths within a system, limiting their maximum mass~\citep{2019MNRAS.485..541C}, and increasing their eccentricities even to the point of migration~\citep{2017AJ....153...42L}, which in turn can have serious consequences on their potential for habitability. We evaluated the coplanarity of the orbits by calculating the 3D angle between the angular momentum vectors of the planetary orbits. {
The unit vectors are given by:
\begin{eqnarray}
    \hat{\bf L}_b =& \sin \Omega_b \sin i_b\,\hat{\bf x} -\cos\Omega_b \sin i_b \,\hat{\bf y} + \cos i_b\,\hat{\bf z}\\
    \hat{\bf L}_c =& \sin \Omega_c \sin i_c\,\hat{\bf x} -\cos\Omega_c \sin i_c \,\hat{\bf y} + \cos i_c\,\hat{\bf  z}
\end{eqnarray}
where $\Omega$ is the longitude of the ascending node and $i$ is the inclination of the orbit. The angle between the two angular momentum vectors is then
\begin{equation}\label{eq:Lbc}
    \Theta_{bc} = \mathrm{cos^{-1}} \left(\hat{\bf L}_b \cdot \hat{\bf L}_c\right)
\end{equation}
Calculating this expression on each chain of our MCMC sampling returns an angle of $\Theta_{bc} = $\thetabc\,degrees. \orvara~places uniform priors on the orientation of each orbital plane, and this translates into a $\sin i$ prior on the relative inclination between the two planes. Comparing the distribution of mutual inclinations to the $\sin i$ prior used in \orvara~to give equal probability to all orbital orientations (Figure~\ref{fig:Lbc}), we see that mutual inclinations of roughly $0-60^{\circ}$ and $130-180^{\circ}$ are strongly disfavored, indicating that the orbits are misaligned. 14~Her joins the ranks of only 3 other systems with giant planets in misaligned orbits: $\pi$ Mensae~\citep{2020A&A...640A..73D}, $\nu$ Andromedae~\citep{2010ApJ...715.1203M}, and Kepler-108~\citep{2017AJ....153...45M}. In comparison with $\pi$~Men~c, whose inclination is not determined and only a plausible range is estimated, we have obtained both inclinations for the 14~Her planets and arrive at a misalignment constraint that is at least as confident as that for $\pi$~Men. 

Another key measurement to constrain the orientation of the system is the spin-axis alignment of the star. Discrepant $V \sin i$ measurements for 14~Her are reported in the literature~\citep{2019AJ....158..101M,2016AandA...596A..76S,2005ApJS..159..141V,2017AJ....153...21L}, giving rise to all range of possible inclinations. However, the brightness of 14~Her makes it potentially amenable for 20\,s-cadence \emph{TESS} observations for asteroseismology to find the projected obliquity of the star (e.g. $\pi$ Men,~\citealt{2021arXiv210809109H}; $\alpha$ Men,~\citealt{2020arXiv201210797C}). A third constraint on the relative angle between the star and the planets would be needed for a full orientation characterization and more clear description of the system's past history.


\begin{figure}
    \centering
    \includegraphics[width=0.77\textwidth]{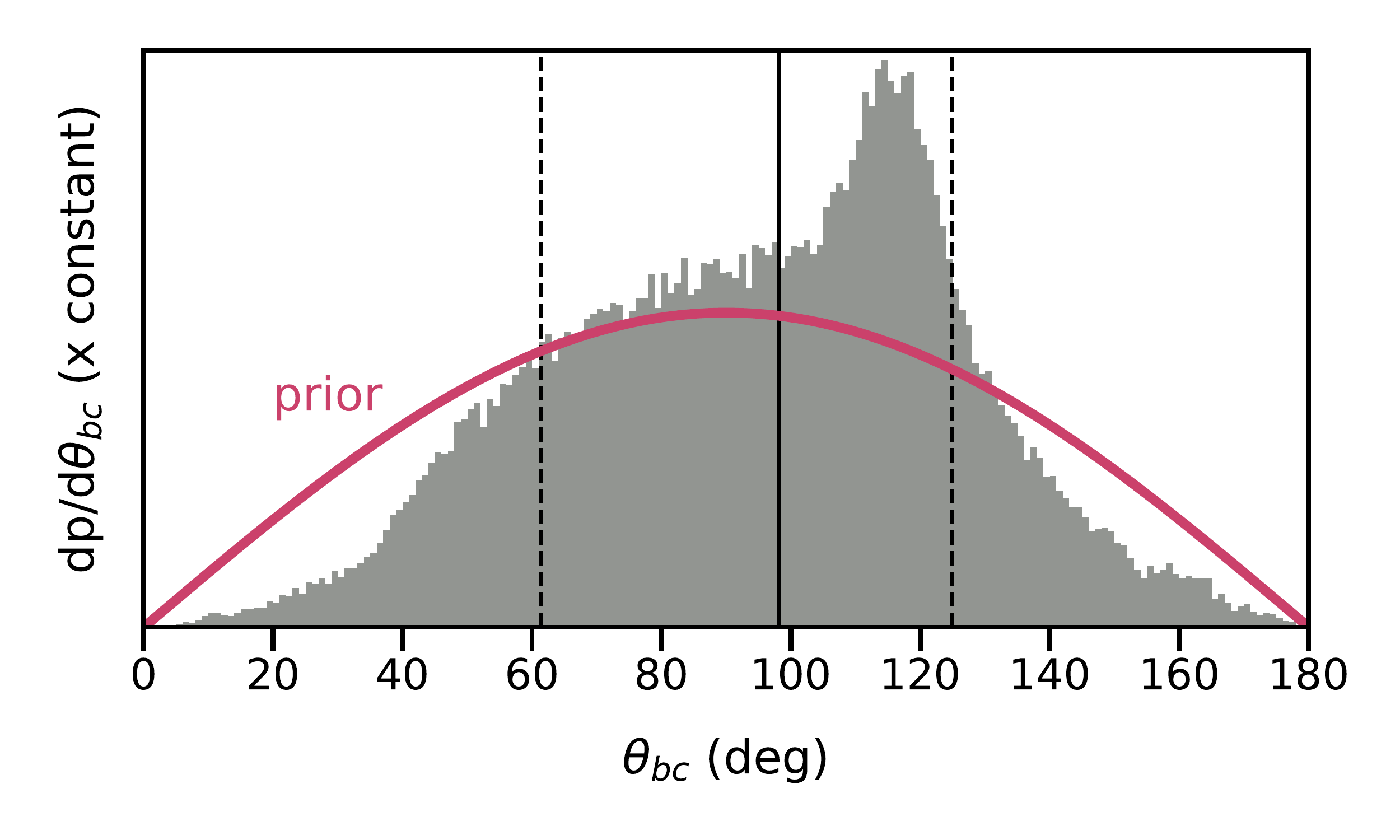}
    \caption{Mutual inclination distribution built from the inclination and ascending node posteriors as shown in Equation~\ref{eq:Lbc}. The the median, $16^{\rm th}$ and $84^{\rm th}$ percentiles are shown as the solid and dashed black lines. While still a broad distribution, the discrepancy between the mutual inclination  and the $\sin i$ prior (shown in maroon) is a strong evidence for the misalignment of the b and c orbits.\label{fig:Lbc}}
\end{figure}

\subsection{Radial velocity offset in HIRES between pre-and post-upgrade}\label{sec:rvoff}

The stability of an RV instrument is crucial to detect small variations that could signal the presence of low-mass and long-period companions. The HIRES spectrograph underwent a major upgrade in 2004 which enabled an RV precision of 1-3\,m/s, a factor of 3 improvement~\citep{2006ApJ...646..505B}. This upgrade changed the RV zero-point by $-1.5\pm0.1\,$m/s, introduced a long-term drift of $<1\,$m/s, and a small nightly drift~\citep{2019MNRAS.484L...8T}.~\citet{2021ApJS..255....8R} modeled the pre- and post-upgrade RV data separately for their planet search without applying the~\citet{2019MNRAS.484L...8T} offset manually. For our results, we conservatively used the~\citet{2021ApJS..255....8R} RV data in the same way, without any modifications. We labeled the pre- and post-upgrade data points as if they were coming from two different instruments to let \orvara~find suitable zero-points and jitter values independently (see Table~\ref{tab:props}). Comparing the zero-points before and after the upgrade, we find a difference of $\Delta_{ZP} = $\rvoff\,m/s, which is larger than the one presented by~\citep{2019MNRAS.484L...8T}.

We tested whether manually adding the~\citet{2019MNRAS.484L...8T} offset to the post-upgrade data would make a difference in our results. We ran four 600,000-step chains and combined them onto a long pseudo-chain with the same settings as for our science results but with a modified RV file with the~\citet{2019MNRAS.484L...8T} offset and two instrument IDs to reflect the pre- and post-upgrade data. The results for both the b and c planet are essentially the same as our science results, including similar jitter values and zero-points. We also tried using the~\citet{2021ApJS..255....8R} data as-is with a single instrument ID. In this case, we found that the posterior parameters for the b planet were the same as our science results within uncertainties, given that the orbital period of the b planet is fully covered several times within the post-upgrade epoch range alone. The median of the posterior parameters of the c planet differ slightly from our science results, especially the semi-major axis ($a_c = {31.1}_{-8.8}^{+16}$\,AU), the eccentricity ($e_c = {0.69}_{-0.11}^{+0.10}$), and the mass ($M_c = {7.11}_{-0.82}^{+1.5}$\,\mjup), although they stayed within the uncertainties of our science results. These differences imply that a small RV offset could have a significant effect in fitting long period planets. We also found more chains caught up in low probability areas compared to our science results. Therefore, it seems that results are robust as long as the pre- and post-upgrade data are treated separately. 


\section{Discussion\label{sec:discussion}}

\subsection{On the formation and evolution of the 14 Her system\label{sec:alignment}}

In today's configuration, 14~Her is a stable system. A planetary system is stable against the mutual gravity of its components if their orbital separations exceed $2\sqrt{3}$ times their mutual Hill radii~\citep{1993Icar..106..247G}, and the large separation between the b and c planets exceeds the stability criterion by a factor of $\sim3$. However, we postulate that today's orbits are likely a far departure from their initial conditions.

Whether resulting from core accretion or gravitational instability, planets are expected to form in disks, causing multi-planet systems to be coplanar as a consequence~\citep{2010fee..book..101H,2010fee..book...71M}. Multi-planet systems tend to have lower eccentricities than single-planet systems~\citep{2009ApJ...693.1084W}, possibly because low eccentricites are energetically favorable for long-term stability. The strongly disfavored coplanarity and high eccentricities in the case of the 14~Her system point to subsequent dynamical evolution following the birth of its planets. 

In a statistical analysis of orbital parameters,~\citet{2013ApJ...767L..24D} found that the orbits of giant planets around metal-rich stars were more eccentric than around metal-poor stars. They postulate that metal-rich stars have solid-rich protoplanetary disks that can form more giant planets than metal-poor stars, which could ultimately engage in gravitational interactions. Multiple planets of similar mass, formed close together, and originally in coplanar, circular orbits can gravitationally excite each other's orbits causing them to become eccentric, misaligned, and occasionally ejecting one planet out of the system in a process called planet-planet scattering~\citep{1996Natur.384..619W,2008ApJ...686..580C,2002aste.book..725M,2010ApJ...711..772R}. Given the high metallicity of 14~Her~A, and the similar masses, large eccentricities, and misaligned orbits of 14~Her~b and c, planet-planet scattering is a likely explanation for the current configuration of the system.

An external possibility is that a stellar fly-by may have triggered gravitational interactions between the planets, causing them to scatter into more eccentric orbits. Stellar fly-bys tend to disrupt a system over time scales of a few million to a few hundred million years and could lead to the ejection of one or more planets within 100\,Myr~\citep{2011MNRAS.411..859M,2015A&A...575A..35B}. 



A more intriguing possibility is that the system initially might have had 3 nearly equal mass giant planets in relatively close, circular, coplanar orbits. After their natal disk is depleted of gas, the planets can engage in close encounters that result in excited orbital eccentricities. In the absence of gas drag, the eccentricities are dissipated through collisions, tidal circularization in the proximity of their stars, or dynamical friction by a residual population of planetesimals~\citep{2013ApJ...775...42I}. At moderate impact parameters, a perturber can cause wide scattering rather than a collision, with recoil velocities close to the planets' surface escape speed. At a distance of a few AU away from the star, this kind of perturbation typically leads to one planet escaping the gravitational potential of the star \citep{2008ApJ...686..580C,2008ApJ...686..603J}. In order to reach stability, the planets that survive develop large eccentricities and widely separated semimajor axes~\citep{2008ApJ...686..621F,2019A&A...629L...7C}, like in the case of the 14~Her system.

The mutual gravity of the planets would have caused the scattering of the most massive one to an eccentric, closer-in orbit (i.e., b with \bmass\,\mjup~at \bsma\,AU and $e = $\becc), the intermediate one to an eccentric, far-out orbit (i.e., c with \cmass\,\mjup~at \csma\,AU and $e = $\cecc), and the ejection of the least massive one out of the system. Initially resonant orbits of 3 coplanar planets can become unstable causing planet-planet scattering, leaving behind a two planet-system with a large semimajor axial ratio ($\alpha = a_b/a_c < 0.3$) with mutual inclinations of $\sim30^{\circ}$ and up to $70^{\circ}$~\citep{2011MNRAS.412.2353L}. With a semimajor axial ratio of 0.12 and mutual inclination of $\Theta_{bc} = $\thetabc\,degrees, the orbital parameters of today's 14~Her system certainly fit these criteria.


The ejected planet would have had a mass lower or equal than that for the c planet ($M_c = $\cmass\mjup), and given the age of the primary star (\age\,Gyr), it would have had a temperature of $\lesssim250\,$K. Isolated, planetary-mass objects at these temperatures are routinely identified as Y dwarfs, the coldest class of brown dwarfs of stellar-like origin~\citep[e.g.,][]{2011ApJ...743...50C,2014ApJ...786L..18L,2019ApJ...881...17M,2020ApJ...895..145B}. Depending on the relative occurrence of planet-planet scattering, the temperature-defined ``Y dwarf'' population might be of mixed origin, with both stellar-born objects and ejected planets among their ranks. 


\subsection{Potential as a future extreme-AO target}

Past direct imaging campaigns have rejected the presence of stellar or substellar companions to 14~Her with confidence. In an effort to identify companions to planet-hosting FGK stars~\citet{2002ApJ...566.1132L} and~\citet{2002ApJ...581..654P} imaged 14~Her as part of their target lists with Keck and Lick adaptive optics (AO), respectively. These studies ruled out the presence of companions up to a magnitude difference of $\Delta K \geq 6.4$\,mag beyond $0\farcs7$ or 12.6\,AU, roughly equivalent to $\geq0.08\,$\msun.~\citet{2009AJ....137..218C} further rejected $K_s = 18$\,mag companions at a $5\farcs0$ separation with Palomar AO.~\citet{2011ApJ...732...10R} used the MMT AO system for deep imaging of 14~Her in the $L'$-band as part of a larger survey to set direct imaging constraints on radial velocity planets.

However, no previous imaging has been able to resolve either planet due to their intrinsic faintness and large contrast with their host star. Based on our derived masses and the age of the star, we estimated effective temperatures and contrasts for the 14~Her planets for a suite of cloudless, hot start evolutionary models~\citep{2008ApJ...689.1327S,2003AandA...402..701B}. We estimate a \teff $= 290-300$\,K for the b planet and \teff $= 260$\,K for the c planet. These cold temperatures in turn imply extreme faintness in NIR bands, leading to contrasts of the order of $10^{-9}-10^{-10}$, far beyond the capabilities of current instrumentation (Table~\ref{tab:contrasts}). Therefore, it is no surprise that these planets have eluded direct detection to date. 

We also estimated the reflected light fraction for both planets by calculating the fraction of light emitted by the star that is intersected by the planet at a given distance and then reflected, based on its global atmospheric properties and phase:

\begin{equation}
    f_R\,(\alpha) = 0.25 \left(\frac{R_p}{a_p}\right)^2 A_B~ \Phi(\alpha)
\end{equation}

where $R_p = 1\,R_\mathrm{Jup}$ is the radius of the planet, $a_p$ is the semimajor axis of the planet, $A_B$ is the Bond albedo of the planet, which we approximate as Jupiter's value~($A_B = 0.503\pm0.012$;~\citealt{2018NatCo...9.3709L}), and $\Phi(\alpha)$ is the Lambert phase function at a given angle $\alpha$~\citep{2012ApJ...747...25M}. The angle $\alpha$ is defined as the angle between the observer, planet, and star with its vertex on the planet. For simplicity, we assume an achromatic phase. We evaluated the reflected fractions when the planets were at quadrature ($\alpha = \pi/2$)
\citep{2012ApJ...747...25M}.

The \teff~of these planets rival some of the coldest known brown dwarfs (e.g., WISE J0830+2837,~\citealt{2020ApJ...895..145B}; WISE J0855$-$0714,~\citealt{2014ApJ...786L..18L}), which are brightest in the mid-infrared~\citep{2012ApJ...756..172M,2011ApJ...743...50C}. Upcoming facilities such as the Near Infrared Camera (NIRCam) aboard the \emph{James Webb Space Telescope (JWST)} are ideally suited to detect objects of these temperatures at high sensitivity, although the contrast with the starlight may impact the detection of planets like 14~Her b or c (Table~\ref{tab:contrasts}). While the b planet has a reflected contrast ratio in the order of $10^{-9}$ and could be potentially detectable with the \emph{Nancy Grace Roman Space Telescope}, its angular proximity to the star ($a_c = 158.9^{+2.1}_{-2.2}$\,mas) could prove challenging for the Coronagraph Instrument. The c planet has a reflected contrast ratio in the order of $10^{-11}$, so even despite its larger angular separation from the star ($a_c = 1529_{-442}^{+869}$\,mas), it is too faint to be detected in optical wavelengths. 

\begin{deluxetable*}{cccccccccc}
\tablenum{3}
\tablecaption{Estimated contrasts and planetary parameters from cloudless evolutionary models.\label{tab:contrasts}}
\tablewidth{0pt}
\tablehead{
\colhead{Component} & 
\colhead{\teff}	& 
\colhead{log $g$} & 
\colhead{L} & 
\colhead{R} & 
\colhead{$J$} & 
\colhead{$K$} & 
\colhead{$f_R$} & 
\colhead{$J$ contrast} & \colhead{$K$ contrast} \\
 & \colhead{(K)} & & (L$_{\odot})$ & (R$_{\odot})$ & (mag) & (mag) &  &  & }
\startdata
star & 	5282 & 4.46	 & $0.67 \pm 0.02$ & $0.99 \pm 0.02$ &	5.158 &	4.714 & \nodata & \nodata & \nodata\\
\hline
\multicolumn{9}{c}{\emph{Sonora-Bobcat}}\\
b & 300	& 4.33	& -7.07	& 0.11 &	27.202 &	24.912 & 1.08E-09	& 2.59E-09 & 9.42E-09 \\
c &	260	& 4.20 & -7.30	& 0.11  & 28.959 & 25.177 & 1.16E-11 & 3.13E-10 & 6.54E-09\\
\hline
\multicolumn{9}{c}{\emph{Saumon \& Marley 2008}}\\
b & 290	& 4.32	& -7.16	& 0.11	& 27.931 &	25.021 & 1.08E-09 & 1.86E-09 & 8.62E-09\\
c & \nodata	& \nodata	& \nodata	& \nodata	& \nodata &	\nodata & 1.16E-11  & \nodata &	\nodata\\
\hline
\multicolumn{9}{c}{\emph{Baraffe et al. 2003}}\\
b &	300	& 4.36	& -7.11	& 0.10	& 27.257 &	24.919	& 1.08E-09 & 2.53E-09 & 9.36E-09\\
c & 260 & 4.22 & -7.34 & 0.10 & 29.013	& 25.185 & 1.16E-11 & 2.99E-10 & 6.49E-09
\enddata
\end{deluxetable*}


\section{Conclusions\label{sec:conclusions}}

In this paper we have characterized the orbital parameters and dynamical evolution of the 14~Her planetary system. Using \orvara, which combines RV and astrometric accelerations, we have obtained a dynamical mass of \bmass\,\mjup~and an inclination of \binc\,degrees for the b planet, hence disentangling the $M\sin i$ degeneracy for this object for the first time. We also set dynamical mass and orbital constraints on the c planet, albeit with larger uncertainties. We have also characterized the fundamental parameters of the star in order to study this system as an ensemble.

Our results describe a middle-aged K0 star with two massive planets in highly eccentric, misaligned orbits. The mutual orientation between the b and c orbits is $\Theta_{bc}$ = \thetabc~degrees. Coplanarity is disfavored for this system, a fact that combined with the large eccentricities, suggests a disruptive planet-planet scattering event leading to the current architecture. An N-body dynamical simulation could strengthen the hypothesis that a third $\lesssim7$\,\mjup~planet was ejected from the system. 

Based on the age of the star and the dynamical planetary masses derived, we infer the effective temperature of the planets from hot start evolutionary models to be 300\,K and 260\,K for b and c, respectively. An ejected planet of these temperatures could be observed today as a planetary-mass Y dwarf.

Future imaging facilities mounted on 30-m class telescopes able to reach NIR contrasts of $10^{-9}$ could potentially be able to directly image these planets for the first time. Based on brown dwarfs studies of objects at similar temperatures~\citep{2014ApJ...793L..16F,2016ApJ...826L..17S,2018ApJ...858...97M}, the best chance of directly imaging the 14~Her planets will be in the mid-infrared. 

\begin{acknowledgements}
We thank the referee and editor for their helpful comments. We thank Sean Raymond for fruitful dynamical discussion about planet-planet scattering. This work has made use of data from the European Space Agency (ESA) mission {\it Gaia} (\url{https://www.cosmos.esa.int/gaia}), processed by the {\it Gaia} Data Processing and Analysis Consortium (DPAC, \url{https://www.cosmos.esa.int/web/gaia/dpac/consortium}). Funding for the DPAC has been provided by national institutions, in particular the institutions participating in the {\it Gaia} Multilateral Agreement. This research has made use of the NASA Exoplanet Archive, which is operated by the California Institute of Technology, under contract with the National Aeronautics and Space Administration under the Exoplanet Exploration Program.
\end{acknowledgements}

\vspace{5mm}
\facilities{Keck(HIRES), \textit{Hipparcos}, \textit{Gaia}, Exoplanet Archive}

\software{astropy \citep{2013AandA...558A..33A}, orvara \citep{2021arXiv210511671B,2021ascl.soft05012B}, SPLAT \citep{2017ASInC..14....7B} }



\bibliography{main}{}
\bibliographystyle{aasjournal}

\end{document}